\documentclass[longauth]{aa} 
\usepackage{graphicx}
\usepackage{txfonts}
\usepackage{natbib}
\usepackage[colorlinks,urlcolor=cyan,citecolor=blue,linkcolor=blue]{hyperref}
\usepackage{amssymb,amsmath}
\usepackage{array}
\usepackage{booktabs}

\begin{document}

\title{55\,Cancri\,e's occultation captured with CHEOPS\thanks{This article uses data from CHEOPS programme CH\_PR100006.}$^{\rm ,}$\thanks{The raw and detrended photometric time-series data are available in electronic form at the CDS via anonymous ftp to cdsarc.u-strasbg.fr (130.79.128.5) or via http://cdsweb.u-strasbg.fr/ cgi- bin/qcat?J/A+A/}}

\author{B.-O.~Demory\inst{1,2}, 
S.~Sulis\inst{3}, 
E.~Meier Vald\'es\inst{1}, 
L.~Delrez\inst{4,5}, 
A.~Brandeker\inst{6}, 
N.~Billot\inst{7}, 
A.~Fortier\inst{2,1}, 
S.~Hoyer\inst{2}, 
S.~G.~Sousa\inst{8}, 
K.~Heng\inst{1,9}, 
M.~Lendl\inst{6}, 
A.~Krenn\inst{10,7}, 
B.~M.~Morris\inst{1},
J.~A.~Patel\inst{6}, 
Y.~Alibert\inst{2}, 
R.~Alonso\inst{11,12}, 
G.~Anglada\inst{13,14}, 
T.~B\'arczy\inst{15}, 
D.~Barrado\inst{16}, 
S.~C.~C.~Barros\inst{8,17}, 
W.~Baumjohann\inst{10}, 
M.~Beck\inst{7}, 
T.~Beck\inst{2}, 
W.~Benz\inst{2,1}, 
X.~Bonfils\inst{18}, 
C.~Broeg\inst{2,19}, 
M.~Buder\inst{20}, 
J.~Cabrera\inst{21}, 
S.~Charnoz\inst{22}, 
A.~Collier Cameron\inst{23}, 
H.~Cottard\inst{24}, 
Sz.~Csizmadia\inst{21}, 
M.~B.~Davies\inst{25}, 
M.~Deleuil\inst{2}, 
O.~D.~S.~Demangeon\inst{8,17}, 
D.~Ehrenreich\inst{26,27}, 
A.~Erikson\inst{21}, 
L.~Fossati\inst{10}, 
M.~Fridlund\inst{28,29}, 
D.~Gandolfi\inst{30}, 
M.~Gillon\inst{4}, 
M.~G\"udel\inst{31}, 
K.~G.~Isaak\inst{32}, 
L.~L.~Kiss\inst{33,34}, 
J.~Laskar\inst{35}, 
A.~Lecavelier des Etangs\inst{36}, 
C.~Lovis\inst{7}, 
A.~Luntzer\inst{37}, 
D.~Magrin\inst{38}, 
L.~Marafatto\inst{38}, 
P.~F.~L.~Maxted\inst{39}, 
V.~Nascimbeni\inst{38}, 
G.~Olofsson\inst{6}, 
R.~Ottensamer\inst{37}, 
I.~Pagano\inst{40}, 
E.~Pall\'e\inst{11}, 
G.~Peter\inst{20}, 
G.~Piotto\inst{38,41}, 
D.~Pollacco\inst{9}, 
D.~Queloz\inst{42,43}, 
R.~Ragazzoni\inst{38,41}, 
N.~Rando\inst{44}, 
F.~Ratti\inst{44}, 
H.~Rauer\inst{21,45,46}, 
I.~Ribas\inst{13,14}, 
N.~C.~Santos\inst{8,17}, 
G.~Scandariato\inst{40}, 
D.~S\'egransan\inst{7}, 
A.~E.~Simon\inst{2}, 
A.~M.~S.~Smith\inst{21}, 
M.~Steller\inst{10}, 
Gy.~M.~Szab\'o\inst{47,48}, 
N.~Thomas\inst{2}, 
S.~Udry\inst{7}, 
V.~Van Grootel\inst{5}, 
N.~A.~Walton\inst{49}}

\institute{\label{inst:1} Center for Space and Habitability, University of Bern, Gesellschaftsstrasse 6, 3012 Bern, Switzerland \and
\label{inst:2} Physikalisches Institut, University of Bern, Sidlerstrasse 5, 3012 Bern, Switzerland \and
\label{inst:3} Aix Marseille Univ, CNRS, CNES, LAM, 38 rue Fr\'ed\'eric Joliot-Curie, 13388 Marseille, France \and
\label{inst:4} Astrobiology Research Unit, Universit\'e de Li\`ege, All\'ee du 6 Ao\^ut 19C, B-4000 Li\`ege, Belgium \and
\label{inst:5} Space sciences, Technologies and Astrophysics Research (STAR) Institute, Universit\'e de Li\`ege, All\'ee du 6 Ao\^ut 19C, 4000 Li\`ege, Belgium \and
\label{inst:6} Department of Astronomy, Stockholm University, AlbaNova University Center, 10691 Stockholm, Sweden \and
\label{inst:7} Observatoire Astronomique de l'Universit\'e de Gen\`eve, Chemin Pegasi 51, CH-1290 Versoix, Switzerland \and
\label{inst:8} Instituto de Astrofisica e Ciencias do Espaco, Universidade do Porto, CAUP, Rua das Estrelas, 4150-762 Porto, Portugal \and
\label{inst:9} Department of Physics, University of Warwick, Gibbet Hill Road, Coventry CV4 7AL, United Kingdom \and
\label{inst:10} Space Research Institute, Austrian Academy of Sciences, Schmiedlstrasse 6, A-8042 Graz, Austria \and
\label{inst:11} Instituto de Astrofisica de Canarias, 38200 La Laguna, Tenerife, Spain \and
\label{inst:12} Departamento de Astrofisica, Universidad de La Laguna, 38206 La Laguna, Tenerife, Spain \and
\label{inst:13} Institut de Ciencies de l'Espai (ICE, CSIC), Campus UAB, Can Magrans s/n, 08193 Bellaterra, Spain \and
\label{inst:14} Institut d'Estudis Espacials de Catalunya (IEEC), 08034 Barcelona, Spain \and
\label{inst:15} Admatis, 5. Kando Kalman Street, 3534 Miskolc, Hungary \and
\label{inst:16} Depto. de Astrofisica, Centro de Astrobiologia (CSIC-INTA), ESAC campus, 28692 Villanueva de la Canada (Madrid), Spain \and
\label{inst:17} Departamento de Fisica e Astronomia, Faculdade de Ciencias, Universidade do Porto, Rua do Campo Alegre, 4169-007 Porto, Portugal \and
\label{inst:18} Universit\'e Grenoble Alpes, CNRS, IPAG, 38000 Grenoble, France \and
\label{inst:19} Center for Space and Habitability, Gesellsschaftstrasse 6, 3012 Bern, Switzerland \and
\label{inst:20} Institute of Optical Sensor Systems, German Aerospace Center (DLR), Rutherfordstrasse 2, 12489 Berlin, Germany \and
\label{inst:21} Institute of Planetary Research, German Aerospace Center (DLR), Rutherfordstrasse 2, 12489 Berlin, Germany \and
\label{inst:22} Universit\'e de Paris, Institut de physique du globe de Paris, CNRS, F-75005 Paris, France \and
\label{inst:23} Centre for Exoplanet Science, SUPA School of Physics and Astronomy, University of St Andrews, North Haugh, St Andrews KY16 9SS, UK \and
\label{inst:24} Almatech, Switzerland \and
\label{inst:25} Centre for Mathematical Sciences, Lund University, Box 118, 221 00 Lund, Sweden \and
\label{inst:26} Observatoire Astronomique de l'Universit\'e de Gen\`eve, Chemin Pegasi 51, CH-1290 Versoix, Switzerland \and
\label{inst:27} Centre Vie dans l'Univers, Facult\'e des sciences, Universit\'e de Gen\`eve, Quai Ernest-Ansermet 30, CH-1211 Gen\`eve 4, Switzerland \and
\label{inst:28} Leiden Observatory, University of Leiden, PO Box 9513, 2300 RA Leiden, The Netherlands \and
\label{inst:29} Department of Space, Earth and Environment, Chalmers University of Technology, Onsala Space Observatory, 439 92 Onsala, Sweden \and
\label{inst:30} Dipartimento di Fisica, Universita degli Studi di Torino, via Pietro Giuria 1, I-10125, Torino, Italy \and
\label{inst:31} University of Vienna, Department of Astrophysics, T\"urkenschanzstrasse 17, 1180 Vienna, Austria \and
\label{inst:32} Science and Operations Department - Science Division (SCI-SC), Directorate of Science, European Space Agency (ESA), European Space Research and Technology Centre (ESTEC),
Keplerlaan 1, 2201-AZ Noordwijk, The Netherlands \and
\label{inst:33} Konkoly Observatory, Research Centre for Astronomy and Earth Sciences, 1121 Budapest, Konkoly Thege Miklos ut 15-17, Hungary \and
\label{inst:34} ELTE E\"otv\"os Lor\'and University, Institute of Physics, P\'azm\'any P\'eter s\'et\'any 1/A, 1117 \and
\label{inst:35} IMCCE, UMR8028 CNRS, Observatoire de Paris, PSL Univ., Sorbonne Univ., 77 av. Denfert-Rochereau, 75014 Paris, France \and
\label{inst:36} Institut d'astrophysique de Paris, UMR7095 CNRS, Universit\'e Pierre et Marie Curie, 98 bis blvd. Arago, 75014 Paris, France \and
\label{inst:37} Department of Astrophysics, University of Vienna, Tuerkenschanzstrasse 17, 1180 Vienna, Austria \and
\label{inst:38} INAF, Osservatorio Astronomico di Padova, Vicolo dell'Osservatorio 5, 35122 Padova, Italy \and
\label{inst:39} Astrophysics Group, Keele University, Staffordshire, ST5 5BG, United Kingdom \and
\label{inst:40} INAF, Osservatorio Astrofisico di Catania, Via S. Sofia 78, 95123 Catania, Italy \and
\label{inst:41} Dipartimento di Fisica e Astronomia "Galileo Galilei", Universita degli Studi di Padova, Vicolo dell'Osservatorio 3, 35122 Padova, Italy \and
\label{inst:42} ETH Zurich, Department of Physics, Wolfgang-Pauli-Strasse 2, CH-8093 Zurich, Switzerland \and
\label{inst:43} Cavendish Laboratory, JJ Thomson Avenue, Cambridge CB3 0HE, UK \and
\label{inst:44} ESTEC, European Space Agency, 2201AZ, Noordwijk, NL \and
\label{inst:45} Zentrum f\"ur Astronomie und Astrophysik, Technische Universit\"at Berlin, Hardenbergstr. 36, D-10623 Berlin, Germany \and
\label{inst:46} Institut f\"ur Geologische Wissenschaften, Freie Universit\"at Berlin, 12249 Berlin, Germany \and
\label{inst:47} ELTE E\"otv\"os Lor\'and University, Gothard Astrophysical Observatory, 9700 Szombathely, Szent Imre h. u. 112, Hungary \and
\label{inst:48} MTA-ELTE Exoplanet Research Group, 9700 Szombathely, Szent Imre h. u. 112, Hungary \and
\label{inst:49} Institute of Astronomy, University of Cambridge, Madingley Road, Cambridge, CB3 0HA, United Kingdom.}

\titlerunning{55\,Cancri\,e's occultation captured with CHEOPS}\authorrunning{B.-O. Demory et al.}

\date{Received}

\abstract{
Past occultation and phase-curve observations of the ultra-short period super-Earth 55\,Cnc\,e obtained at visible and infrared wavelengths have been challenging to reconcile with a planetary reflection and emission model. In this study, we analyse a set of 41 occultations obtained over a two-year timespan with the CHEOPS satellite. We report the detection of 55\,Cnc\,e's occultation with an average depth of $12\pm3$ ppm. We derive a corresponding 2-$\sigma$ upper limit on the geometric albedo of $A_g < 0.55$ once decontaminated from the thermal emission measured by \textit{Spitzer} at 4.5$\mu$m. CHEOPS's photometric performance enables, for the first time, the detection of individual occultations of this super-Earth in the visible and identifies short-timescale photometric corrugations likely induced by stellar granulation. We also find a clear 47.3-day sinusoidal pattern in the time-dependent occultation depths that we are unable to relate to stellar noise, nor instrumental systematics, but whose planetary origin could be tested with upcoming JWST occultation observations of this iconic super-Earth.}


\maketitle

\section{Introduction}

A striking feature in the population of exoplanets is the paucity of short-period objects ($P\lesssim 1$ day) with sizes larger than 2 $R_{\oplus}$. The vast majority of these ultra-short period (USP) planets have sizes comparable to that of the Earth \citep{Sanchis-Ojeda:2014,Wu:2013,Owen:2013,Fulton:2017a}. We may thus wonder whether these planets were formerly gas or ice giants that experienced dramatic erosion \citep[e.g.][]{Baraffe:2005, Jackson:2013, Owen:2013, Lundkvist:2016, Jin:2018, Lee:2018}. \citet{Winn:2017} provided convincing evidence that USP planets are likely not eroded hot-Jupiters \citep{Lammer:2009}, based on the stark difference in metallicity between the hot-Jupiter and USP-planet host stars. It is still debated whether USP planets are remnants of smaller, sub-Neptune-sized planets or rocky planets that did not contain a significant amount of volatiles at the time of formation \citep{Venturini:2020}. In this context, the characterisation of USP planet atmospheres is of prime importance \citep{Raymond:2018}.

The super-Earth 55\,Cnc\,e stands out as an attractive target to understand the population of USP exoplanets. With a radius of $1.88\pm0.03$ $R_{\oplus}$, a mass of $8.0\pm0.3$ $M_{\oplus}$ \citep{Dawson:2010, Winn:2011a, Demory:2011, Bourrier:2018a}, and an orbital period of only 17.7 hr, 55\,Cnc\,e is larger than most USP planets. 55\,Cnc\,e's bulk density favours the presence of an atmosphere \citep{Dorn:2017, Crida:2018, Bourrier:2018a}, possibly SiO- \citep{Schaefer:2009} or N-dominated \citep{Miguel:2019}, but so far no observations have provided definitive evidence for such an atmosphere \citep{Ehrenreich:2012, Ridden-Harper:2016,Tsiaras:2016,Esteves:2017,Jindal:2020,Tabernero:2020,Deibert:2021,Zhang:2021,Keles:2022}.

55\,Cnc\,e's previous long-stare continuous observations in the visible have complicated the picture further by showing planetary phase-curve amplitudes $>80$ ppm, largely exceeding the thermal+reflected contribution from the planet, with no detectable occultations \citep{Winn:2011a, Dragomir:2014, Sulis:2019,Morris:2021a}. This observation shows that, contrary to other USP planets such as LHS\,3844\,b \citep{Vanderspek:2019,Kreidberg:2019} or K2-141\,b \citep{Zieba:2022}, which can be accurately modelled with bare rock surfaces, visible observations of 55\,Cnc\,e require a more complex interplay of processes, possibly involving  the star \citep{Morris:2021a} or even circumstellar material \citep{Gillon:2012}. While 55\,Cnc\,e's \textit{MOST} visible phase-curves display varying amplitudes and offsets with time, there is only weak evidence for variability in the occultation depths measured with \textit{TESS} on inter-sector timescales \citep{Meier-Valdes:2022}.

To investigate the unusual behaviour of 55\,Cnc\,e at visible wavelengths, we conducted a large programme in the frame of the CHEOPS Guaranteed Time Observations comprising 777 hours. In this paper we investigate the origin of the occultation signal from a set of 41 eclipses and discuss the implications for 55\,Cnc\,e's geometric albedo.

\section{Methods}\label{sec:obs}

\subsection{CHEOPS observations and data reduction}\label{sec:reduc}

CHEOPS \citep{Benz:2021a} observed 55\,Cnc in the frame of the Guaranteed Time Observations Programme CH\_PR100006. The observations acquired between March 2020 and February 2022 consist of 29 visits in total, each comprising 16 CHEOPS orbits (except 25 orbits for visits \#28 and \#29), with an approximate 100-minute duration each. While this study focuses on 55\,Cnc\,e's occultations, a separate paper will detail the phase-curve analysis (Meier-Vald\'es et al., in prep.). Because of the Sun-synchronous, dawn-to-dusk orbit of CHEOPS combined to the +28$^{\circ}$ declination of 55\,Cnc, the time spent on source was mostly limited by Earth occultations, South-Atlantic anomaly interruptions, and stray light contamination, leading to a typical duty cycle between 53 and 61\%. Individual frames (integration time of 2.2 s) were co-added on board into 44.2 s stacks, which corresponds to the data cadence, to reduce the downlink volume. A summary of all CHEOPS time-series photometric observations of 55\,Cnc, out of which the occultations were extracted, is shown in Table~\ref{tab:log}.

\begin{table*}
\begin{center}
\caption{Log of CHEOPS time-series photometric observations of 55\,Cnc, out of which we extracted the occultations of 55\,Cnc\,e.\label{tab:log}}
\begin{tabular}{l c c c c c}
\toprule
Visit \# & Start time [MJD] &  End time [MJD] &  File key & Frame count & Efficiency [\%] \\
\midrule
1  &    58931.5730  &   58932.6668  &   CH\_PR100041\_TG000601\_V0200  &        1213  &      56.7 \\
2  &    59184.5902  &   59185.7266  &   CH\_PR100006\_TG000301\_V0200  &        1178  &      53.0 \\
3  &    59200.2144  &   59201.2841  &   CH\_PR100006\_TG000302\_V0200  &        1171  &      56.0 \\
4  &    59201.3138  &   59202.4393  &   CH\_PR100006\_TG000303\_V0200  &        1189  &      54.0 \\
5  &    59207.7710  &   59208.8402  &   CH\_PR100006\_TG000304\_V0200  &        1194  &      57.1 \\
6  &    59208.8693  &   59209.9395  &   CH\_PR100006\_TG000305\_V0200  &        1200  &      57.4 \\
7  &    59210.8231  &   59211.9311  &   CH\_PR100006\_TG000306\_V0200  &        1226  &      56.6 \\
8  &    59224.2610  &   59225.3281  &   CH\_PR100006\_TG000307\_V0200  &        1244  &      59.6 \\
9  &    59225.3638  &   59226.4272  &   CH\_PR100006\_TG000308\_V0200  &        1237  &      59.5 \\
10 &    59228.4152  &   59229.5097  &   CH\_PR100006\_TG000309\_V0200  &        1233  &      57.6 \\
11 &    59229.5172  &   59230.6118  &   CH\_PR100006\_TG000310\_V0200  &        1257  &      58.8 \\
12 &    59232.7825  &   59233.8771  &   CH\_PR100006\_TG000311\_V0200  &        1254  &      58.6 \\
13 &    59233.8849  &   59235.0126  &   CH\_PR100006\_TG000312\_V0200  &        1317  &      59.7 \\
14 &    59238.7863  &   59239.8871  &   CH\_PR100006\_TG000313\_V0200  &        1264  &      58.7 \\
15 &    59243.9759  &   59245.0704  &   CH\_PR100006\_TG000314\_V0200  &        1264  &      59.1 \\
16 &    59247.7693  &   59248.8639  &   CH\_PR100006\_TG000315\_V0200  &        1266  &      59.2 \\
17 &    59250.4992  &   59251.5673  &   CH\_PR100006\_TG000316\_V0200  &        1270  &      60.8 \\
18 &    59267.3978  &   59268.4654  &   CH\_PR100006\_TG000601\_V0200  &        1269  &      60.8 \\
19 &    59276.0585  &   59277.1531  &   CH\_PR100006\_TG000602\_V0200  &        1248  &      58.3 \\
20 &    59279.5099  &   59280.6045  &   CH\_PR100006\_TG000603\_V0200  &        1241  &      58.0 \\
21 &    59283.2943  &   59284.4034  &   CH\_PR100006\_TG000701\_V0200  &        1231  &      56.8 \\
22 &    59294.7374  &   59295.8055  &   CH\_PR100006\_TG000702\_V0200  &        1191  &      57.0 \\
23 &    59573.4394  &   59574.6002  &   CH\_PR100006\_TG000901\_V0200  &        1257  &      55.4 \\
24 &    59591.3659  &   59592.4299  &   CH\_PR100006\_TG000401\_V0200  &        1246  &      59.9 \\
25 &    59592.9374  &   59594.0091  &   CH\_PR100006\_TG000402\_V0200  &        1263  &      60.3 \\
26 &    59594.0360  &   59595.1652  &   CH\_PR100006\_TG000403\_V0200  &        1333  &      60.4 \\
27 &    59595.1728  &   59596.2673  &   CH\_PR100006\_TG000404\_V0200  &        1253  &      58.6 \\
28 &    59629.6846  &   59631.3758  &   CH\_PR100006\_TG001201\_V0200  &        1998  &      60.4 \\
29 &    59636.6283  &   59638.3134  &   CH\_PR100006\_TG001301\_V0200  &        1939  &      58.9 \\

\hline 
\end{tabular}
\end{center}
\end{table*}


The CHEOPS sub-array (200$\times$200 pixels) data were first processed using the mission's data reduction pipeline (DRP, version 13.1.0) described in \citet{Hoyer:2020}, which is based on aperture photometry. For all visits, we used the \texttt{DEFAULT} aperture of 25 pixels, which yields the smallest photometric root mean square (RMS) according to the DRP report. We then complemented the DRP output with point spread function (PSF) photometry using the PSF Imagette Photometric Extraction (\texttt{PIPE})\footnote{\url{https://github.com/alphapsa/PIPE}} pipeline, which is a Python package developed for CHEOPS (Brandeker et al. in prep). A useful benefit of \texttt{PIPE} is that it allows us to quantify the PSF scale dependence on the spacecraft orbital phase. In addition, we employed PIPE to process the 30-pixel radii imagettes that were extracted from the sub-arrays, stacked in pairs on board providing a 2.2-s cadence. A visual inspection of the full-array frames shows smearing from the neighbouring 53\,Cnc (V=6.23), which contaminates 55\,Cnc's (V=5.98) aperture on CHEOPS orbital timescales. The sub-array DRP and PIPE outputs, as well as the PIPE imagette reductions, provide final time series that are corrected from the smearing of 53\,Cnc. 

For each reduction, we then removed data points flagged by the pipeline because of, for example, cosmic rays or stray light contamination and mask background, and $X$-$Y$ centroid position outliers identified by large median absolute deviations from the median, assuming a Normal distribution and using a consistent estimator of the standard deviation being $\hat\sigma = 1.4826 \times {\rm MAD}$ \citep{Rousseeuw:1993}. This final step of the data reduction resulted in about 12\% of the frames being discarded for the analysis, out of which an average of 5\% were due to bad quality flag from the DRP and 7\% due to the large background. We attribute this larger number \citep[compared to other CHEOPS studies, e.g.][]{Brandeker:2022} to the close contaminant and the sky location of 55\,Cnc leading to more severe stray light contamination. We finally compared the DRP and PIPE sub-array reductions and found that the latter yield a slightly better flux RMS (82 vs 76 ppm), and we therefore retained the PIPE sub-array photometry for the subsequent analysis.

\subsection{Data analysis}\label{sec:analysis}

In this section we describe how the occultation depths were derived from the reduced CHEOPS time series. We first cut the 29 visits into individual data segments with a single occultation each. We defined a data segment as valid if its duration was at least three times the occultation duration, to ensure that an accurate instrumental baseline model could be derived from the out-of-eclipse photometry. This step resulted in a sample of 41 individual occultations. 

For the treatment of CHEOPS instrumental systematics, we employed a similar approach to the marginalisation method presented by \citet{Gibson:2014a} and used in the past for other facilities \citep[e.g. HST,][]{Wakeford:2016}. The main reason for this approach is that CHEOPS orbital period (100 min), and therefore its associated systematics, is close to 55\,Cnc\,e's occultation duration ($\sim$95 min). Marginalising over the instrumental models reduces the incidence of a particular baseline model biasing the retrieved parameters. For this purpose, we used a modified version of the MCMC algorithm implementation already presented in the literature \citep[e.g.][]{Gillon:2012a,Demory:2012,Gillon:2014a}. The inputs to the MCMC fit are the CHEOPS photometric time series described in Sect.~\ref{sec:reduc}. We performed an initial fit to measure the noise properties in our time series. We estimated the underestimation or overestimation of white noise ($\beta_w$), which is the ratio of the un-binned residual RMS over the photometric error. We also computed an estimate for the correlated noise ($\beta_r$), which is the ratio of the binned RMS over Gaussian scaling of the photometric error on the same timescale \citep[e.g.][]{Winn:2008b, Delrez:2018}. We computed $\beta_r$ over 10- to 120-minute timescales and adopted the most conservative value. 

For each light curve, we used 16 different baseline models that we fitted simultaneously to the occultation depth. The photometric baseline model coefficients used for CHEOPS systematics detrending were determined at each step of the MCMC procedure using a singular value decomposition method \citep{Press:1992}. The resulting coefficients were then used to correct the raw photometric light curves. Our 16 baseline models encompassed a sum of polynomial functional forms of the roll angle (first and second order), time (up to first order), centroid position in $X$ and $Y$ (up to first order), and background (up to first order). For each model, we approximated the Bayesian evidence using the Bayesian Information Criterion \citep[BIC,][]{Schwarz:1978} and, rather than selecting the model with the best BIC, we computed the marginalised occultation parameters from the three baseline models yielding the smallest BIC values. We then multiplied the time series uncertainties by the average correction factor ${CF}= \beta_w \times \beta_r$ determined from the three selected baseline models to account for the white- and red-noise estimates from the initial fit.

Our second fit was conducted in a global fashion, where the only free parameter was the occultation depth, where negative values were allowed in the fit to prevent an overestimation of the shallow signals \citep[e.g.][]{Demory:2014}. Because of the strong correlated noise in the data (Sect.~\ref{sec:var}), we chose to hold the transit parameters (transit centre, period, impact parameter, and $a/R_{\star}$) fixed (see Table~\ref{tab:params}) to avoid exploring unrealistic combinations of the multidimensional parameter space, which would happen if we used priors on these parameters \citep[e.g.][]{Garhart:2020}. We also assumed circular orbits and fixed $\sqrt{e}\cos \omega$ and $\sqrt{e} \sin \omega$ values to 0 \citep{Nelson:2014a}. We ran two chains of 100\,000 steps (including 20\% burn-in) each, and checked their efficient mixing and convergence by visually inspecting the autocorrelation functions of each chain, and by using the \citet{Gelman:1992} statistical test, ensuring that the test values for all fitted parameters were $<1.01$.
Our third fit was identical to the preceding one, with the exception that each individual occultation was fit independently of the rest of the dataset.
Finally, we repeated this entire analysis using PIPE's imagette photometry to assess whether the higher cadence impacted our occultation depth median values and corresponding credible intervals.

\begin{table}
\begin{center}
\caption{List of 55\,Cnc\,e parameters values used in our global analysis.\label{tab:params}}
\begin{tabular}{l c l}
\toprule
Parameter (fixed) & Value & Reference \\
\midrule
Transit centre $T_0$    &    $ 2458932.00042$   & \citet{Morris:2021a}   \\
Orbital period $P$ [d]  &    $ 0.73654737$      & \citet{Bourrier:2018a} \\
Impact parameter $b$    &    $ 0.39$            & \citet{Bourrier:2018a} \\
$\sqrt{e}\cos \omega$   &    $ 0.0 $              & \citet{Nelson:2014a} \\
$\sqrt{e}\sin \omega$   &    $ 0.0 $              & \citet{Nelson:2014a} \\
\midrule
Parameter (fitted) & Value & Reference \\
\midrule
$dF_{\rm occ}$ [ppm] & $12\pm3$ & This work \\
\hline 
\end{tabular}
\end{center}
\end{table}

\section{Results}\label{sec:results}

\begin{figure}[!ht]
    \centering
    \includegraphics[width=0.48\textwidth]{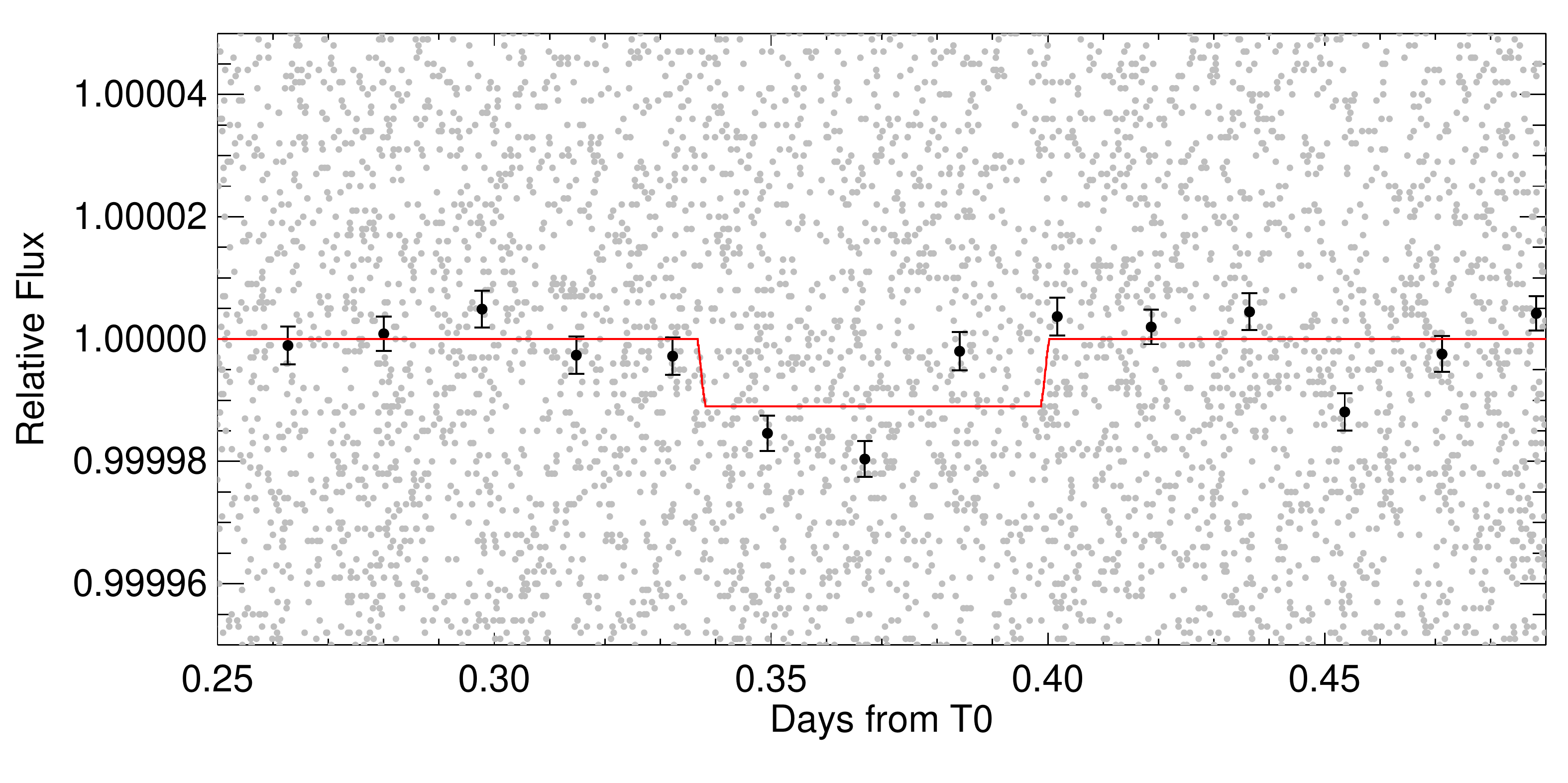}
    \caption{CHEOPS phase-folded occultation of 55\,Cnc\,e from the global analysis. Grey points are un-binned data points while 15-minute bins are shown as black points. The best-fit model is shown in red. The deviations from the occultation model that are not accounted for by the instrument baseline model suggest a source of correlated noise from planetary or stellar origin.}
    \label{fig:lc}
\end{figure}

The global fit, including all 41 orbits (2020 to 2022), yielded an occultation depth of $dF_{\rm occ} = 12\pm3$ ppm (see Fig.~\ref{fig:lc}) with $ {\rm BIC}=10891$, compared to ${\rm BIC}=10896$ without the occultation model included. We thus assess the $\Delta {\rm BIC}=5$ as moderate evidence for the detection of 55\,Cnc\,e's occultation in the averaged dataset. We show all individual occultation depths from our third fit (Sect.~\ref{sec:analysis}) in Table~\ref{tab:depths} and they are also depicted in Fig.~\ref{fig:depths}. We note that using sub-arrays or imagettes in the analysis yield results (parameter values and credible intervals) that are indistinguishable from each other. We thus elected to conduct the analysis using the sub-array frames for computational efficiency.

We used the same method as in \citet{Meier-Valdes:2022} to compute the corresponding geometric albedo in the CHEOPS bandpass, using a \texttt{PHOENIX} stellar spectrum \citep{Husser:2013} with $T_{\rm eff}=5200$\,K, [Fe/H]=0.3, and $\log g=4.5$ \citep{von-Braun:2011a}. We assumed a blackbody planetary emission spectral energy distribution. We find that at face value, the $12\pm3$ ppm occultation depth translates into a brightness temperature  of $T_{B} = 2843^{+93}_{-114} K$ in the CHEOPS bandpass or into a geometric albedo of $A_g = 0.42\pm0.12$ if it were reflected light alone. A significant contribution to the CHEOPS $T_B$ originates from the planetary thermal emission, for which a maximum dayside-hemisphere brightness temperature of $T_B = 2705^{+249}_{-252}$\,K was measured with \textit{Spitzer} at 4.5$\mu$m \citep{Tamburo:2018b}. This thermal contamination translates into 4 to 15 ppm (8 ppm for the mean $T_B = 2705$\,K) in the CHEOPS bandpass.\footnote{Compared to 11 ppm for the mean $T_B = 2705$\,K in the redder TESS bandpass \citep{Meier-Valdes:2022,Kipping:2020b}.}

The reflected light contribution in the CHEOPS bandpass results in a 55\,Cnc\,e decontaminated geometric albedo $A_g = 0.15^{+0.20}_{-0.19}$ (2-$\sigma$ upper limit of $A_g < 0.55$). Assuming that 55\,Cnc\,e has no volatile envelope, its \textit{Spitzer} 4.5$\mu$m brightness temperature would point to a high-albedo CaO-Al$_{\rm 2}$O$_{\rm 3}$ molten surface with $A_g \sim 0.5$ \citep{Rouan:2011,Kite:2016,Essack:2020}. Assuming a thick atmosphere scenario for 55\,Cnc\,e yields a wide range of geometric albedos that strongly depend on the atmospheric composition and the possible formation of clouds \citep{Hammond:2017, Angelo:2017, Mahapatra:2017}. The sole knowledge of the geometric albedo of such a hot super-Earth does not enable us to distinguish between a molten surface and a thick atmosphere scenario.

\begin{table}
\begin{center}
\caption{Measured occultation timings (BJD$_{\rm TDB}$ $-$2\,450\,000), depths in ppm with uncertainties, residual RMS (in ppm at a sub-array cadence of 44 s), and white- ($\beta_w$)  and red- ($\beta_r$) noise correction factors (see text). \label{tab:depths}}
\renewcommand{\arraystretch}{1.4}
\begin{tabular}{l l l c l c c c c}
\toprule
\# & Timing  & Depth & RMS & $\beta_w$ & $\beta_r$ \\
\midrule
1  &   8932.10871    &    $ -23_{ -19}^{+19}$   &       69 & 1.36 & 1.48 \\
2  &   8932.79614    &    $  20_{ -33}^{+37}$   &       69 & 1.34 & 2.48 \\
3  &   9185.42750    &    $  -3_{ -16}^{+15}$   &       67 & 1.31 & 1.36 \\
4  &   9200.93497    &    $  45_{ -12}^{+12}$   &       62 & 1.22 & 1.00 \\
5  &   9201.63159    &    $ 166_{ -26}^{+22}$   &       75 & 1.47 & 1.90 \\
6  &   9202.37023    &    $ -12_{ -17}^{+17}$   &       68 & 1.33 & 1.37 \\
7  &   9208.27724    &    $ -15_{ -12}^{+13}$   &       69 & 1.35 & 1.03 \\
8  &   9208.99714    &    $ -48_{ -21}^{+24}$   &       70 & 1.36 & 1.77 \\
9  &   9209.73371    &    $  33_{ -13}^{+14}$   &       67 & 1.30 & 1.00 \\
10 &   9211.94330    &    $  -2_{ -20}^{+18}$   &       74 & 1.45 & 1.42 \\
11 &   9225.20107    &    $ -11_{ -16}^{+16}$   &       84 & 1.65 & 1.04 \\
12 &   9225.93799    &    $ -21_{ -16}^{+16}$   &       83 & 1.63 & 1.06 \\
13 &   9226.68685    &    $ -22_{ -22}^{+20}$   &       73 & 1.41 & 1.46 \\
14 &   9229.63948    &    $  63_{ -18}^{+19}$   &       77 & 1.50 & 1.36 \\
15 &   9230.35707    &    $   3_{ -16}^{+17}$   &       78 & 1.52 & 1.28 \\
16 &   9233.30309    &    $  17_{ -13}^{+13}$   &       79 & 1.54 & 1.00 \\
17 &   9234.03952    &    $  23_{ -13}^{+11}$   &       78 & 1.53 & 1.00 \\
18 &   9234.79574    &    $  -8_{ -16}^{+16}$   &       75 & 1.46 & 1.00 \\
19 &   9239.94868    &    $  21_{ -18}^{+16}$   &       81 & 1.58 & 1.17 \\
20 &   9245.09376    &    $  48_{ -15}^{+14}$   &       79 & 1.55 & 1.00 \\
21 &   9248.77071    &    $  23_{ -14}^{+14}$   &       79 & 1.55 & 1.00 \\
22 &   9251.00527    &    $  58_{ -14}^{+15}$   &       78 & 1.52 & 1.06 \\
23 &   9251.71684    &    $  16_{ -24}^{+26}$   &       87 & 1.69 & 1.61 \\
24 &   9267.92099    &    $  10_{ -25}^{+23}$   &       78 & 1.51 & 1.70 \\
25 &   9268.65738    &    $ -18_{ -22}^{+28}$   &       79 & 1.52 & 1.93 \\
26 &   9276.76142    &    $ -32_{ -14}^{+13}$   &       73 & 1.43 & 1.00 \\
27 &   9280.47084    &    $   6_{ -29}^{+25}$   &       75 & 1.45 & 2.02 \\
28 &   9284.12510    &    $   7_{ -16}^{+18}$   &       80 & 1.57 & 1.14 \\
29 &   9295.24081    &    $  53_{ -18}^{+18}$   &       73 & 1.43 & 1.00 \\
30 &   9295.92594    &    $  16_{ -28}^{+29}$   &       76 & 1.47 & 2.20 \\
31 &   9574.34935    &    $  12_{ -16}^{+17}$   &       71 & 1.38 & 1.18 \\
32 &   9592.00641    &    $ -47_{ -16}^{+16}$   &       89 & 1.74 & 1.00 \\
33 &   9593.47491    &    $   9_{ -15}^{+15}$   &       83 & 1.61 & 1.00 \\
34 &   9594.21135    &    $  11_{ -15}^{+14}$   &       85 & 1.66 & 1.00 \\
35 &   9594.95962    &    $   2_{ -22}^{+21}$   &       79 & 1.53 & 1.42 \\
36 &   9595.71194    &    $  15_{ -30}^{+28}$   &       84 & 1.63 & 1.62 \\
37 &   9596.42095    &    $  12_{ -19}^{+23}$   &       89 & 1.73 & 1.00 \\
38 &   9630.32797    &    $  57_{ -17}^{+14}$   &       85 & 1.64 & 1.00 \\
39 &   9631.03901    &    $  13_{ -13}^{+14}$   &       71 & 1.37 & 1.12 \\
40 &   9637.67818    &    $ -41_{ -16}^{+15}$   &       75 & 1.46 & 1.02 \\
41 &   9638.43423    &    $   4_{ -14}^{+13}$   &       74 & 1.44 & 1.02 \\
\hline 
\end{tabular}
\renewcommand{\arraystretch}{1.0}
\end{center}
\end{table}

\section{Discussion}\label{sec:var}

\begin{figure*}
    \centering
    \includegraphics[width=\textwidth]{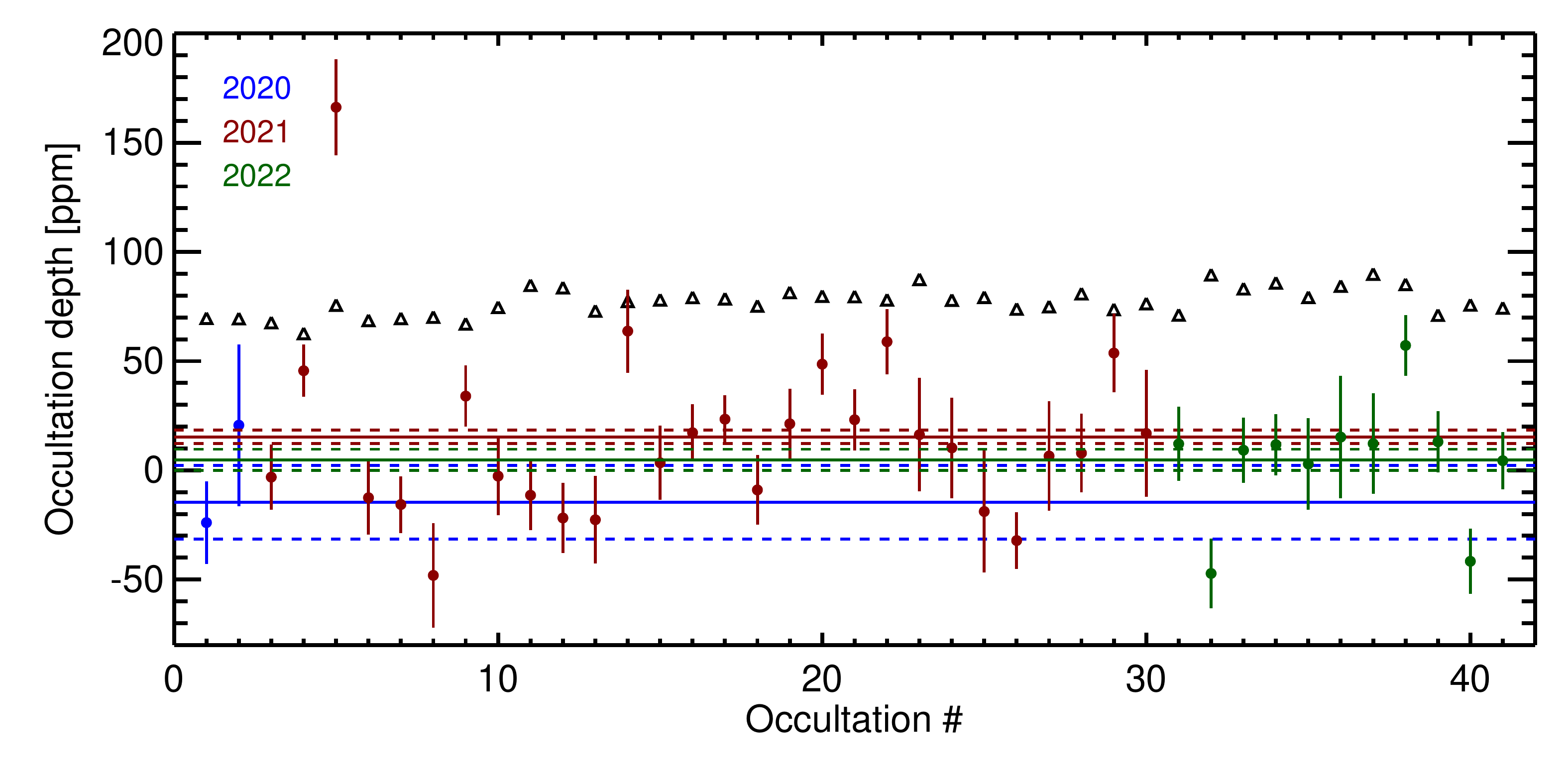}
    \caption{Individual occultation depths of 55\,Cnc\,e from the global analysis. Occultation epochs are related to the timings shown in Table~\ref{tab:depths}. Each year is colour-coded, where data obtained in 2020, 2021, and 2022 are are shown in blue, red, and green, respectively (see also Appendix~\ref{app:occdepths}). Corresponding weighted means (plain lines) and credible intervals (dashed lines) are also shown. The residual photometric RMS for each occultation are shown as black triangles.}
    \label{fig:depths}
\end{figure*}

Two caveats jeopardise the accuracy of the parameters derived in the global fit and the inferred CHEOPS geometric albedo. First, our phase-folded light curve (Fig.~\ref{fig:lc}) shows significant correlated noise, but we do not find significant inter-seasonal variations (see  Fig.~\ref{fig:depths} and Appendix~\ref{app:occdepths}), consistent with \citet{Meier-Valdes:2022}. Second, the 4.5 $\mu$m maximum brightness temperature of 2700\,K used above for 55\,Cnc\,e's thermal contribution may not be representative of the planet thermal emission at the time of the observations \citep{Demory:2016,Tamburo:2018b}, which would result in an underestimated reflected light contribution. We discuss these issues in the next two sections.

\begin{figure}[!ht]
    \centering
    \includegraphics[width=0.5\textwidth]{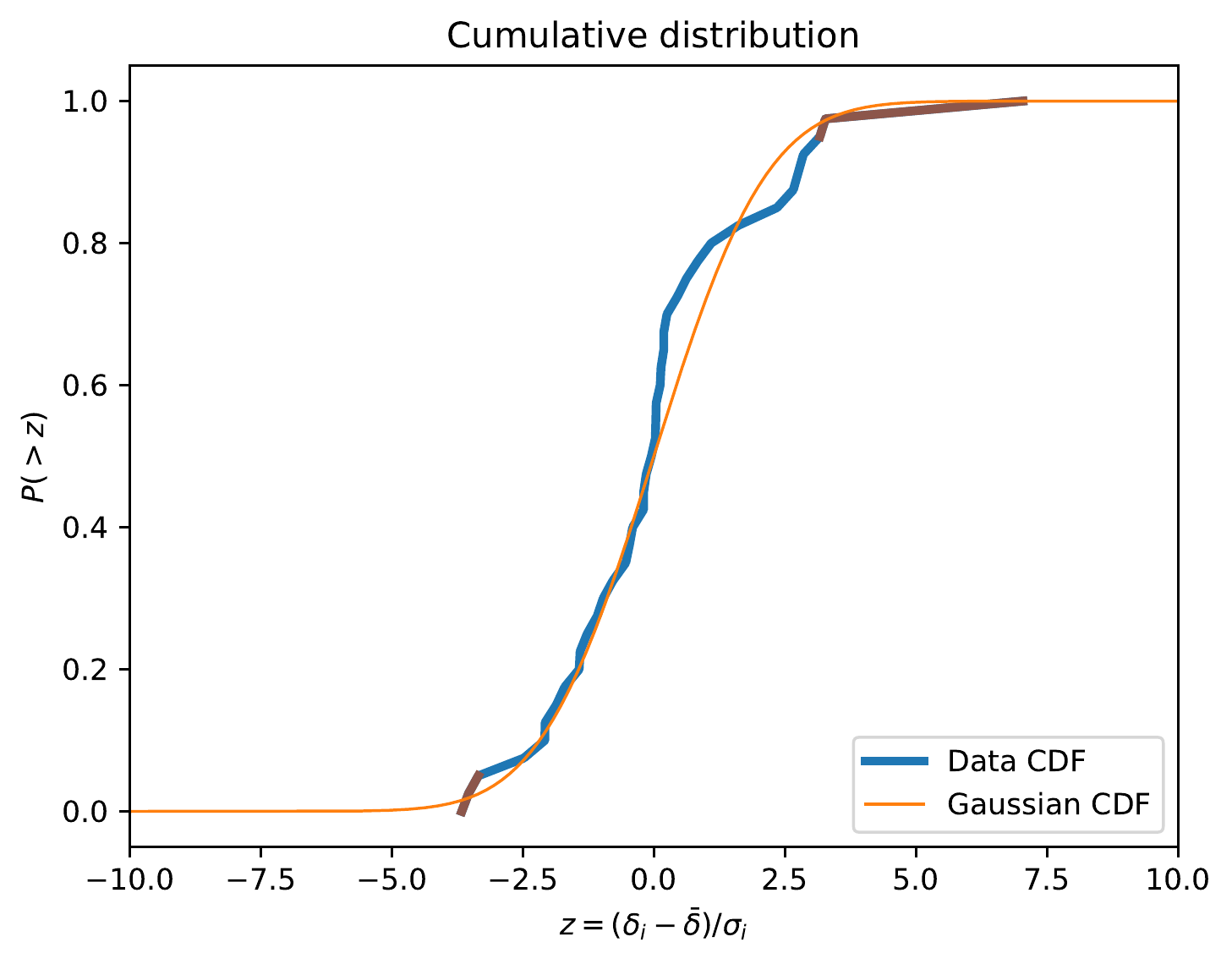}
    \caption{Cumulative PDF of all standardised occultation depths. The CDF corresponding to the CHEOPS data is shown in blue (with 3-$\sigma$ outliers in brown). For comparison, a Gaussian CDF is shown in orange. This figure's code is adapted from Agol et al. (2020).}
    \label{fig:cdf}
\end{figure}

To investigate the origin of the scatter in Fig.~\ref{fig:depths}, we first computed the cumulative distribution function (CDF) of the standardised residuals, where we superimposed the CDF of a Gaussian distribution (Fig.~\ref{fig:cdf}). One limitation is that our sample of 41 occultations was statistically small. We note that the occultation depth's CDF follows a Gaussian behaviour for negative values. This sub-sample, therefore, is Normal-distributed and is not affected by significant correlated noise at zeroth order. The positive values, however, exhibit a higher probability around the mean, and an excess of events at $z \sim 2.5$ (depth $\delta \sim$ 50 ppm) and a single $z>7$ event at $\delta$=166 ppm, where $z = (\delta_{i}-\bar{\delta})/\sigma_{i}$. We show some of these individual occultations in Figure~\ref{fig:seldepths}. 

\begin{figure}[!ht]
    \centering
    \includegraphics[width=0.5\textwidth]{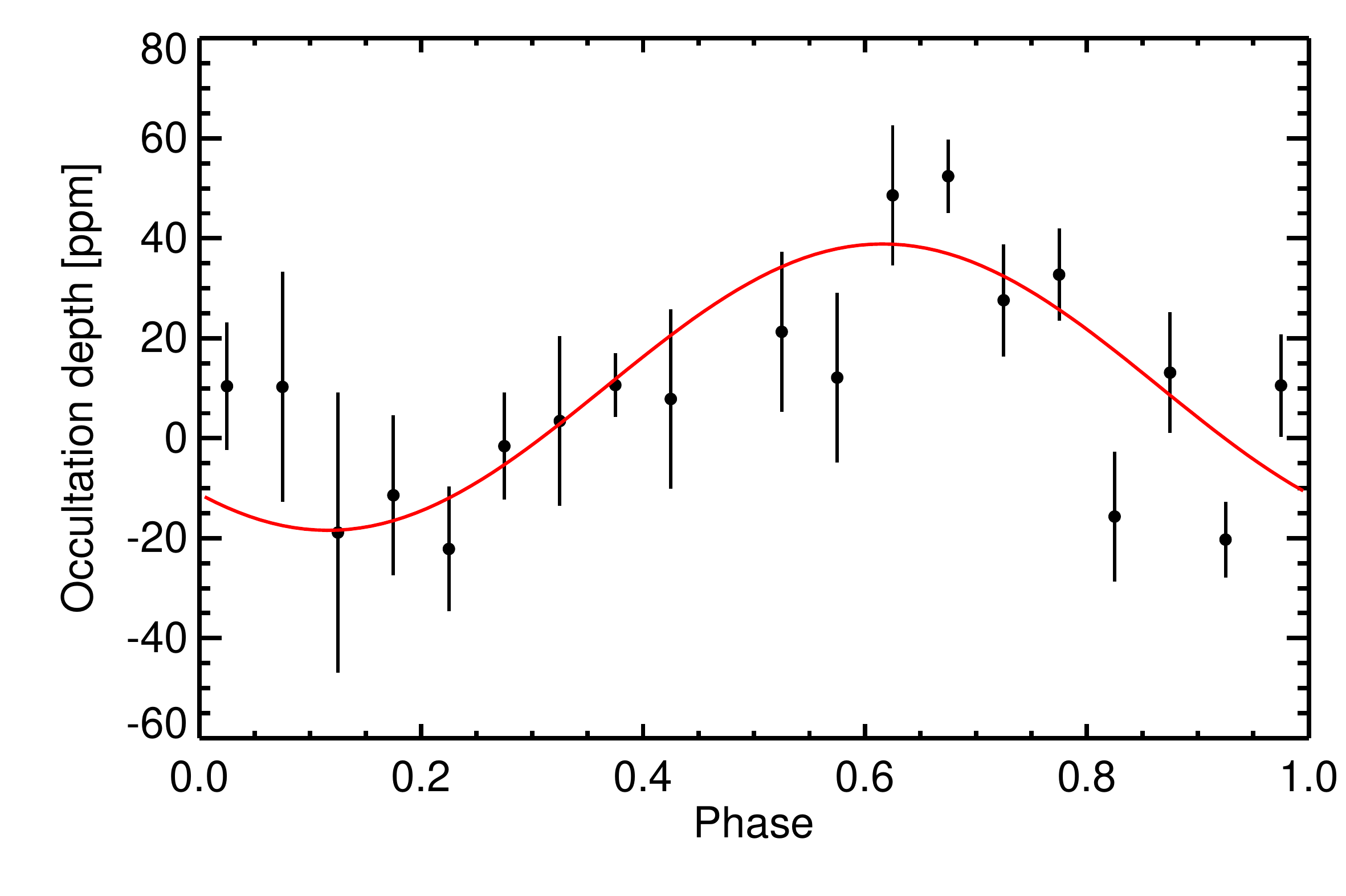}
    \caption{Phase-folded long-term variability of 55\,Cnc\,e's occultation depths. The binned data (20 bins for 41 points) are shown in black, while our best-fit model (sine function with offset) is shown in red. The phase axis corresponds to the P=47.3-day period.}
    \label{fig:occvar}
\end{figure}

Similar to \citet{Tamburo:2018b}, we then attempted to characterise the putative variability timescale and amplitude by employing two different models to reproduce the occultation depth versus time pattern. For each model fitted to the data, we computed the BIC for model comparison purposes. Model 1 is a flat line, whose only free parameter is the value at origin, which yields a BIC= 173. Model 2 is a sine function with an offset of the form $a \sin(2 \pi \omega t + b) + c $, where $\omega$ is the sine function's frequency independently determined prior to the fit, using an adaptative grid with initial periods ranging from 4 to 100 days. We find $\omega$=0.0211 days$^{-1}$ (i.e. a 47.3-day period), $a=29\pm4$ ppm, $b=-127.55\pm0.12$, and $c=10\pm3$ ppm (close to our 12 ppm mean) to yield the smallest BIC = 126. We show the corresponding model fit, along the phase-folded occultation depths, in Fig.~\ref{fig:occvar}. This simple model comparison strongly favours ($\Delta {\rm BIC} = 47$) that the 41 occultations collected by CHEOPS over the past two years are better reproduced by a variable, sinusoidal pattern, than by a constant, flat line.

\subsection{Instrumental origin}
\label{sec:instru}
We investigated whether the CHEOPS spacecraft and instrument could be at the origin of the variability behaviour described above. We repeated the same analysis as the one described in Sect.~\ref{sec:analysis} but on out-of-eclipse time series where there are no occultations, extracted at the same orbital phase across all visits. We ran an MCMC analysis on that dataset, with the only difference compared with Sect.~\ref{sec:analysis} and Table~\ref{tab:params} being a 0.18-day shift in the $T_0$ to match the data given as input to the MCMC. This global fit yielded a depth of -1$\pm$3 ppm, which points to the instrument being innocuous to the occultation signal previously detected at the expected 55\,Cnc\,e phase. We also investigated any dependence of this analysis's measured depths on time, as done above for the occultation dataset. We find that the 47.3-day peak is entirely absent from the periodogram and that the dominant frequency is at 0.22 days$^{-1}$, which is also present in the occultation data and an alias of the window function. We finally note that the photometric residual noise measured in this analysis's and the occultation's time series are in good agreement (76 vs 77 ppm respectively). This analysis discards the instrument as the source of the observed depth variations.

\subsection{Planetary origin}

Given the past evidence of 55\,Cnc\,e's brightness temperature variability measured at 4.5 $\mu$m with \textit{Spitzer} \citep{Demory:2016, Tamburo:2018b}, we explored whether the planet could be at the origin of the variable depth pattern detected in the visible with CHEOPS. The excess of events around occultation depths of $\delta\sim$50 ppm (Fig.~\ref{fig:cdf}) corresponds to a hemisphere-averaged brightness temperature of $\sim$3500\,K, which is significantly above the maximum brightness temperature of 2892$\pm$280\,K measured in the infrared with \textit{Spitzer}. Even assuming such a large thermal contribution would lead to an unrealistically high geometric albedo $A_g\sim1.58$, so these 50 ppm cannot be explained by reflected light alone. We thus do not have evidence that a surge in thermal emission could be at the origin of this excess in CHEOPS brightness temperature. We may hypothesise that the longer-term 47.3-day sinusoidal pattern of the CHEOPS occultation depths has a common origin with the $\sim$35-day sinusoidal variation detected in the eight \textit{Spitzer} occultations acquired between 2012 and 2013 \citep{Tamburo:2018b}. Upcoming JWST occultation observations may help in investigating this hypothesis further. We finally note for completeness that the non-transiting 51.2-$M_{\odot}$ planet 55\,Cnc\,c has an orbital period of $P=44.4$ days, which is close to the period of our sinusoidal model fit. The required reflected flux from 55\,Cnc\,c to contribute to the drastic changes in 55\,Cnc\,e's occultation depth would, however, have to be unrealistically high (see also Sect.~\ref{sec:stellar}).

\subsection{Stellar origin}
\label{sec:stellar}
We then investigated whether the photometric residuals seen in the CHEOPS occultation time series ($\sim$5 hours each) could be explained by stellar granulation noise. Other studies have hinted that the sensitivity of high-precision photometric  observations are due to to the small-amplitude granulation stellar signal, which becomes dominant in the high-signal-to-noise regime of bright stars as seen for \textit{Kepler} \citep{Lally:2022} or CHEOPS \citep[e.g.][]{Delrez:2021,Sulis:2022}, and is thus particularly relevant for 55\,Cnc.

The RMS of the CHEOPS 55\,Cnc occultation time series range between $62$ and $89$ ppm (see Table~\ref{tab:depths}). This is consistent with the photon noise level of $66.2$ ppm that is predicted by the \texttt{CHEOPS Exposure Time Calculator} (ETC)\footnote{\url{https://cheops.unige.ch/pht2/exposure-time-calculator}}  at the cadence of the downloaded images (44.2s) and the efficiency of these observations (53 to 61\%).

\begin{figure}[!ht]
\centering
\includegraphics[width=0.5\textwidth]{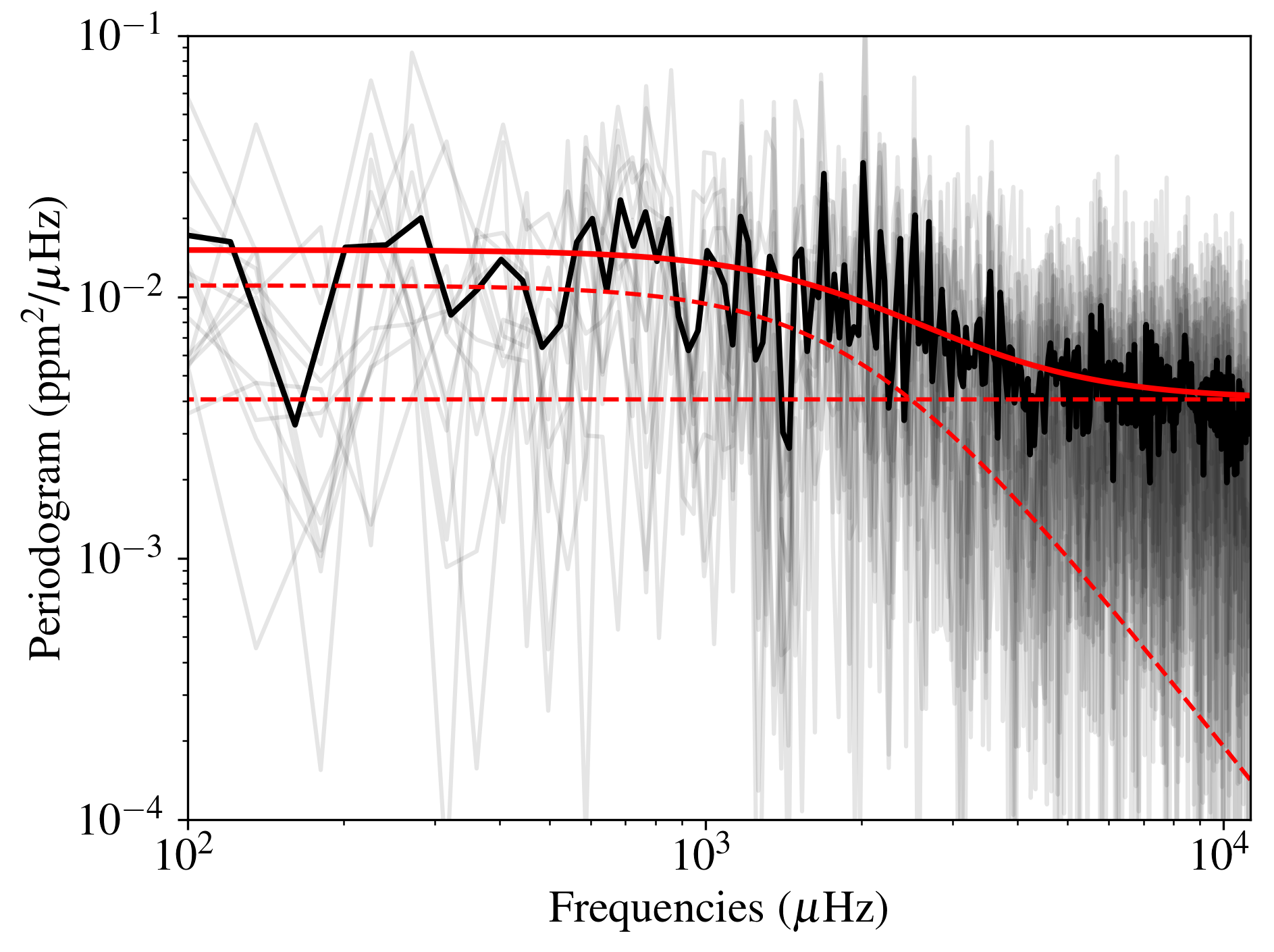}
\caption{\label{fig:periodo}Periodogram of the residuals of each CHEOPS observation (black) and their respective averaged periodogram (thick black). The solid red line represents the best-fitting model given in  Eq.(\ref{eq:harvey}), and the dashed red lines show the two components of this model.}
\end{figure}

The occultation time-series residuals reveal a clear dependence on frequency (Fig.~\ref{fig:periodo}, black). The corresponding periodogram shows an increase in power towards lower frequencies in the $[596, 5365]~\mu$Hz region (i.e. periods $\sim [3, 27]$ min, which matches 55\,Cnc\,e's occultation ingress and egress' duration) that is characteristic of stellar granulation for solar-like stars. However, the lower-frequency end ($1000~\mu$Hz and below) does not exhibit significant correlated noise. We computed the averaged periodogram of the time-series residuals (thick black line) and fitted a classical Harvey function of the form \citep{Kallinger:2014}:

\begin{equation}
P_H(\nu_k^{+}) := \eta^2(\nu_k^{+}) \frac{a_g}{1+\Big(\frac{\nu_k^{+}}{b_g}\Big)^{c_g}} + \sigma^2,
\label{eq:harvey}
\end{equation}

where $\{a_g, b_g, c_g\}$ are the amplitude, characteristic frequency, and power of the stellar granulation signal, and $\sigma^2$ is the variance of the high-frequency noise component, which is dominated by photon noise. The notation $\nu_k^{+}$ means that only positive frequencies are considered, and $\eta(\nu_k^{+}) := {\rm sinc}(\frac{\pi}{2}\frac{\nu_k^{+}}{\nu_{Ny}})$ is an attenuation factor based on the Nyquist frequency. The best-fitting Harvey function is shown in red in Fig.~\ref{fig:periodo}. 

We obtained $a_g= 0.0108$ ppm$^2$/$\mu$Hz, $b_g=1563.5$ $\mu$Hz, $c_g=3.3$ and $\sigma^2=0.0046$ ppm$^2$/$\mu$Hz. The level of the high-frequency noise $\sigma=67.8$ ppm is fully consistent with the ETC prediction. Assuming an oscillation frequency at maximum power of $\nu_{max}= 3323.57$ $\mu$Hz from asteroseismology \citep{Kjeldsen:1995}, the characteristic amplitude $a_g$ and frequency $b_g$ are also found to be fully consistent with predictions from \textit{Kepler} observations \citep[see Fig.~8 of][]{Kallinger:2014}. Finally, 3D hydrodynamical models of stellar granulation predict an RMS of $45.2$ ppm for 55\,Cnc\,A \citep[see Table~2 of][]{Rodriguez-Diaz:2022}. To compare this value with our dataset, we binned the occultation time series into $3$-minute intervals and filtered the periods $>27$ min \citep[see details in][]{Sulis:2022}. We computed an RMS between $36$ and $60$ ppm, which is consistent with these 3D model's predictions.

55\,Cnc's granulation amplitude and timescale thus appears as a plausible cause of the correlated noise observed in the individual occultation time series. Because of the partial coverage of our occultations, granulation could impact only part of each occultation, biasing the individual retrieved depth by several tens of ppm. One issue for this hypothesis, however, is that the shallow and negative occultation depths are Normal-distributed, as discussed in Sect.~\ref{sec:var}, while stellar granulation should induce at zeroth order the same level of  correlated noise throughout all datasets. We note that a straightforward way to improve the quality of our fit would be to model granulation noise using Gaussian processes \citep{Aigrain:2015,Barros:2020}, but the reliability of this extra step would rely on the assumption that the occultation depth is constant across visits, which is challenged by the observed infrared occultation depth variability and the long-term sinusoidal pattern (Sect.~\ref{sec:var}). Since thermal emission accounts for roughly two-thirds of the signal measured in the CHEOPS bandpass, this approach could induce a bias of up to 8 ppm in the retrieved occultation depth for the thermal contribution alone.

We finally note that the sinusoidal model period of 47.3 days found above appears very close to the rotation period of the star, which has been previously reported from spectroscopy to be 38.8$\pm$0.05 days, but with a significant inter-season scatter \citep{Bourrier:2018a}, and from photometry to be 42.7$\pm$2.5 days \citep{Fischer:2008}. Our sinusoidal model fit yields a peak-to-valley change in the occultation depth (not in the stellar flux) of 58 ppm, which cannot be explained by the small, rotational 6-mmag photometric variability reported from 10 years of monitoring from the star with the automatic photometric telescope at Fairborn Observatory in the visible \citep{Fischer:2008}. The sinusoidal variations in the occultation depths are also challenging to explain by granulation, as no dependence of this process is expected on the stellar rotational period. Furthermore, the 47.3-day modulation does not appear in our out-of-eclipse data analysis (Sect.~\ref{sec:instru}), which advocates against a stellar origin.

\section{Conclusions}\label{sec:concl}

CHEOPS detects 55\,Cnc\,e's occultation in the visible, but also points to a large scatter that cannot be accounted for with Gaussian noise alone. The CHEOPS observations suggest that this larger scatter, recently studied with TESS data, can be explained by astrophysical correlated noise and that an instrumental source can be discarded. The power density spectrum of the model residuals shows that the amplitude and timescale of the correlated noise is compatible with stellar granulation, whose amplitude exceeds 55\,Cnc\,e's average occultation depth and whose typical frequencies span 55\,Cnc\,e's occultation ingress and egress duration, which, coupled to the observations' average efficiency biases the individual occultation depths. The variations in occultation depth observed in the visible with CHEOPS are likely affected by this short-timescale stellar noise and are of too small amplitude to be detected with TESS due to the larger photon noise. We also argue that granulation is not the cause for the observed variability in the \textit{Spitzer} bandpass, since only the occultations have been shown to vary, not the transits \citep{Demory:2016,Tamburo:2018b}, and with a significantly larger amplitude than the observed RMS in the CHEOPS observations. Furthermore, granulation noise is of lower amplitude in the infrared than in the visible. 

In summary, while stellar granulation can explain the short-timescale noise seen in our data, it cannot impact the observed longer-timescale variations. We cannot identify a process originating from the instrument, nor from the star that could explain the 47-day modulation of the occultation depths detected with CHEOPS. With the data at hand, we are not able to discard a planetary origin for this signal. Two different JWST programmes will observe 55\,Cnc\,e's occultations in the infrared during cycle 1, using the NIRCam (PID 2084, PI A.~Brandeker) and MIRI (PID 1952, PI. R.~Hu) instruments, which will undoubtedly help us understand better \citep{Zilinskas:2022} the nature of this iconic super-Earth.

\begin{acknowledgements}
We thank the referee, Jacob Bean, and the editor, Emmanuel Lellouch, for a very careful reading of this manuscript and useful suggestions that improved the paper.
CHEOPS is an ESA mission in partnership with Switzerland with important contributions to the payload and the ground segment from Austria, Belgium, France, Germany, Hungary, Italy, Portugal, Spain, Sweden, and the United Kingdom. The CHEOPS Consortium would like to gratefully acknowledge the support received by all the agencies, offices, universities, and industries involved. Their flexibility and willingness to explore new approaches were essential to the success of this mission. 
B.-O.D. acknowledges insightful discussions with Hannah Diamond-Lowe and Jo\~ao Mendon\c ca on the way to Andechs monastery, a lot of help from Patricio Cubillos on bibtex management \citep{Cubillos2020zndoBibmanager} as well as support from the Swiss National Science Foundation (PP00P2\_190080)and the Swiss State Secretariat for Education, Research and Innovation (SERI) under contract number MB22.00046.
The Belgian participation to CHEOPS has been supported by the Belgian Federal Science Policy Office (BELSPO) in the framework of the PRODEX Program, and by the University of Liege through an ARC grant for Concerted Research Actions financed by the Wallonia-Brussels Federation. 
L.D. is an F.R.S.-FNRS Postdoctoral Researcher. 
SH gratefully acknowledges CNES funding through the grant 837319. 
S.G.S. acknowledge support from FCT through FCT contract nr. CEECIND/00826/2018 and POPH/FSE (EC). 
ABr was supported by the SNSA. 
ML acknowledges support of the Swiss National Science Foundation under grant number PCEFP2\_194576. 
YA and MJH acknowledge the support of the Swiss National Fund under grant 200020\_172746. 
We acknowledge support from the Spanish Ministry of Science and Innovation and the European Regional Development Fund through grants ESP2016-80435-C2-1-R, ESP2016-80435-C2-2-R, PGC2018-098153-B-C33, PGC2018-098153-B-C31, ESP2017-87676-C5-1-R, MDM-2017-0737 Unidad de Excelencia Maria de Maeztu-Centro de Astrobiologia (INTA-CSIC), as well as the support of the Generalitat de Catalunya/CERCA programme. The MOC activities have been supported by the ESA contract No. 4000124370. 
S.C.C.B. acknowledges support from FCT through FCT contracts nr. IF/01312/2014/CP1215/CT0004. 
XB, SC, DG, MF and JL acknowledge their role as ESA-appointed CHEOPS science team members. 
ACC acknowledges support from STFC consolidated grant numbers ST/R000824/1 and ST/V000861/1, and UKSA grant number ST/R003203/1. 
This project was supported by the CNES. 
This work was supported by FCT - Fundacao para a Ciencia e a Tecnologia through national funds and by FEDER through COMPETE2020 - Programa Operacional Competitividade e Internacionalizacao by these grants: UID/FIS/04434/2019, UIDB/04434/2020, UIDP/04434/2020, PTDC/FIS-AST/32113/2017 \& POCI-01-0145-FEDER- 032113, PTDC/FIS-AST/28953/2017 \& POCI-01-0145-FEDER-028953, PTDC/FIS-AST/28987/2017 \& POCI-01-0145-FEDER-028987, O.D.S.D. is supported in the form of work contract (DL 57/2016/CP1364/CT0004) funded by national funds through FCT. 
This project has received funding from the European Research Council (ERC) under the European Union's Horizon 2020 research and innovation programme (project {\sc Four Aces}. 
grant agreement No 724427). It has also been carried out in the frame of the National Centre for Competence in Research PlanetS supported by the Swiss National Science Foundation (SNSF). DE acknowledges financial support from the Swiss National Science Foundation for project 200021\_200726. 
MF and CMP gratefully acknowledge the support of the Swedish National Space Agency (DNR 65/19, 174/18). 
DG gratefully acknowledges financial support from the CRT foundation under Grant No. 2018.2323 ``Gaseous or rocky? Unveiling the nature of small worlds''. 
M.G. is an F.R.S.-FNRS Senior Research Associate. 
KGI is the ESA CHEOPS Project Scientist and is responsible for the ESA CHEOPS Guest Observers Programme. She does not participate in, or contribute to, the definition of the Guaranteed Time Programme of the CHEOPS mission through which observations described in this paper have been taken, nor to any aspect of target selection for the programme. 
This work was granted access to the HPC resources of MesoPSL financed by the Region Ile de France and the project Equip@Meso (reference ANR-10-EQPX-29-01) of the programme Investissements d'Avenir supervised by the Agence Nationale pour la Recherche. 
PM acknowledges support from STFC research grant number ST/M001040/1. 
LBo, GBr, VNa, IPa, GPi, RRa, GSc, VSi, and TZi acknowledge support from CHEOPS ASI-INAF agreement n. 2019-29-HH.0. 
This work was also partially supported by a grant from the Simons Foundation (PI Queloz, grant number 327127). 
IRI acknowledges support from the Spanish Ministry of Science and Innovation and the European Regional Development Fund through grant PGC2018-098153-B- C33, as well as the support of the Generalitat de Catalunya/CERCA programme. 
GyMSz acknowledges the support of the Hungarian National Research, Development and Innovation Office (NKFIH) grant K-125015, a a PRODEX Experiment Agreement No. 4000137122, the Lend\"ulet LP2018-7/2021 grant of the Hungarian Academy of Science and the support of the city of Szombathely. 
V.V.G. is an F.R.S-FNRS Research Associate. 
NAW acknowledges UKSA grant ST/R004838/1.

\end{acknowledgements}

\bibliographystyle{aa}
\bibliography{ch_55cnc.bib}

\begin{appendix}

\section{Seasonal 55\,Cnc\,e's occultation light curves}
\label{app:occdepths}

\begin{figure}[!ht]
    \centering
    \includegraphics[width=0.48\textwidth]{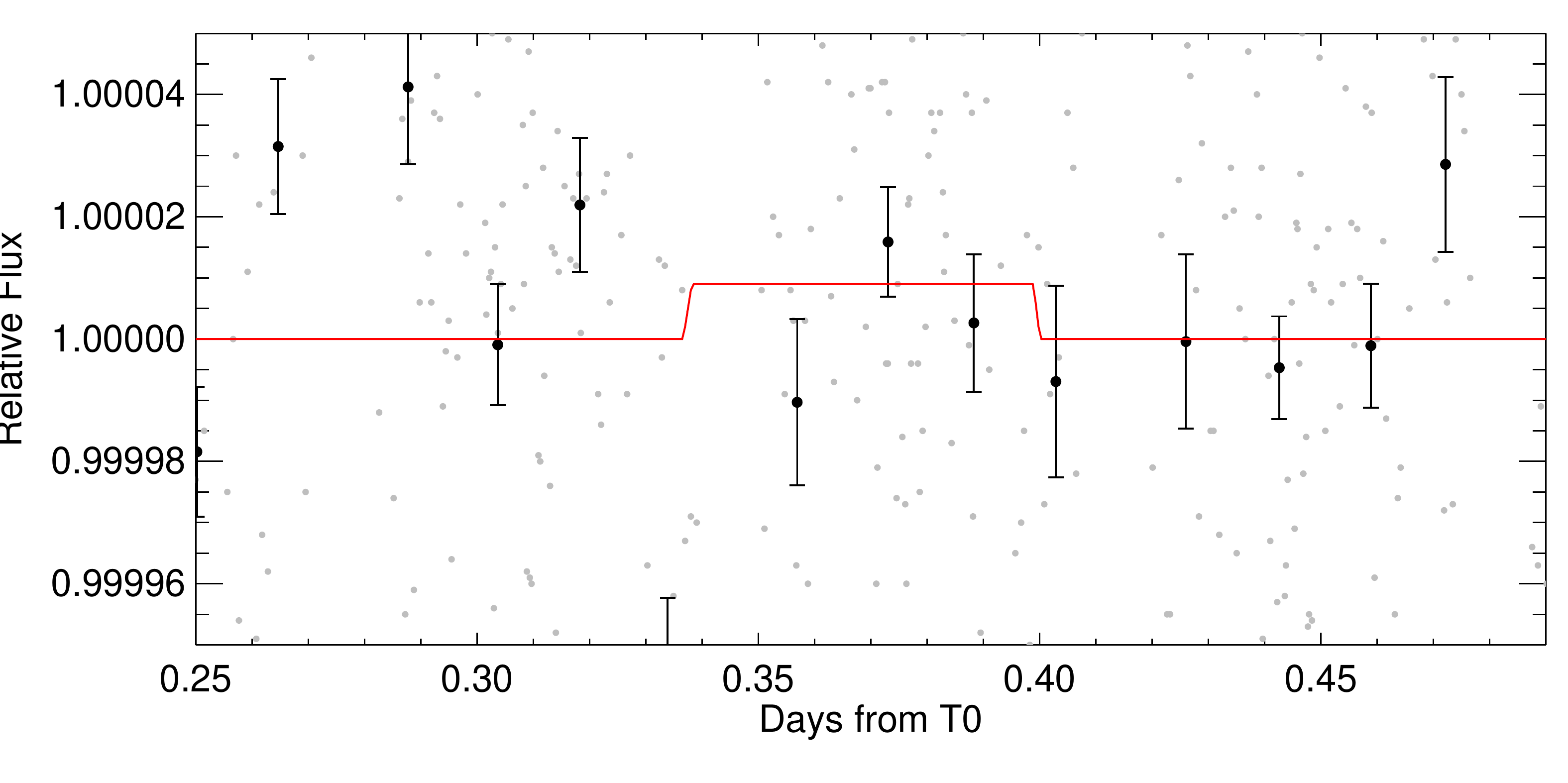}
    \includegraphics[width=0.48\textwidth]{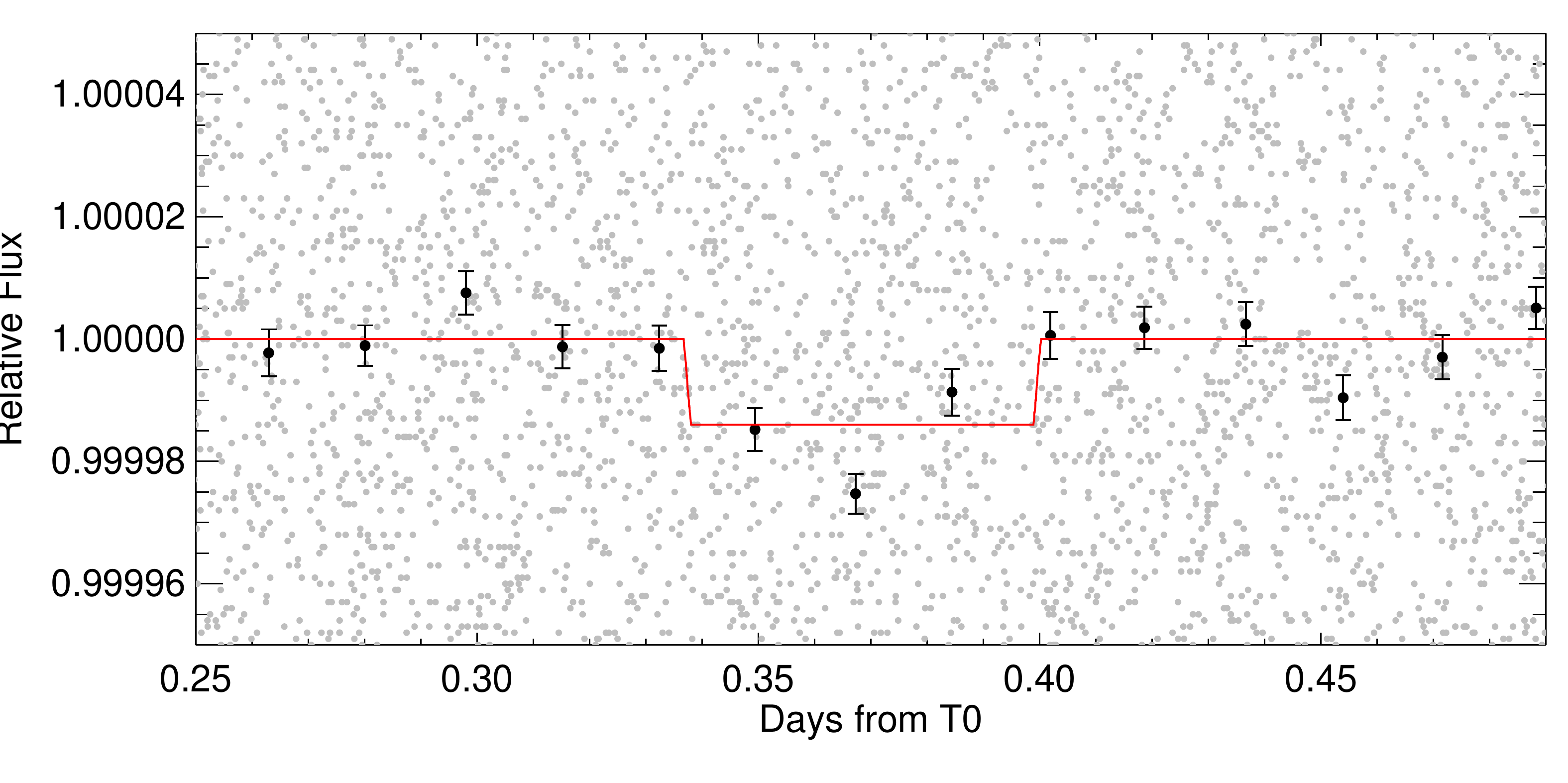}
    \includegraphics[width=0.48\textwidth]{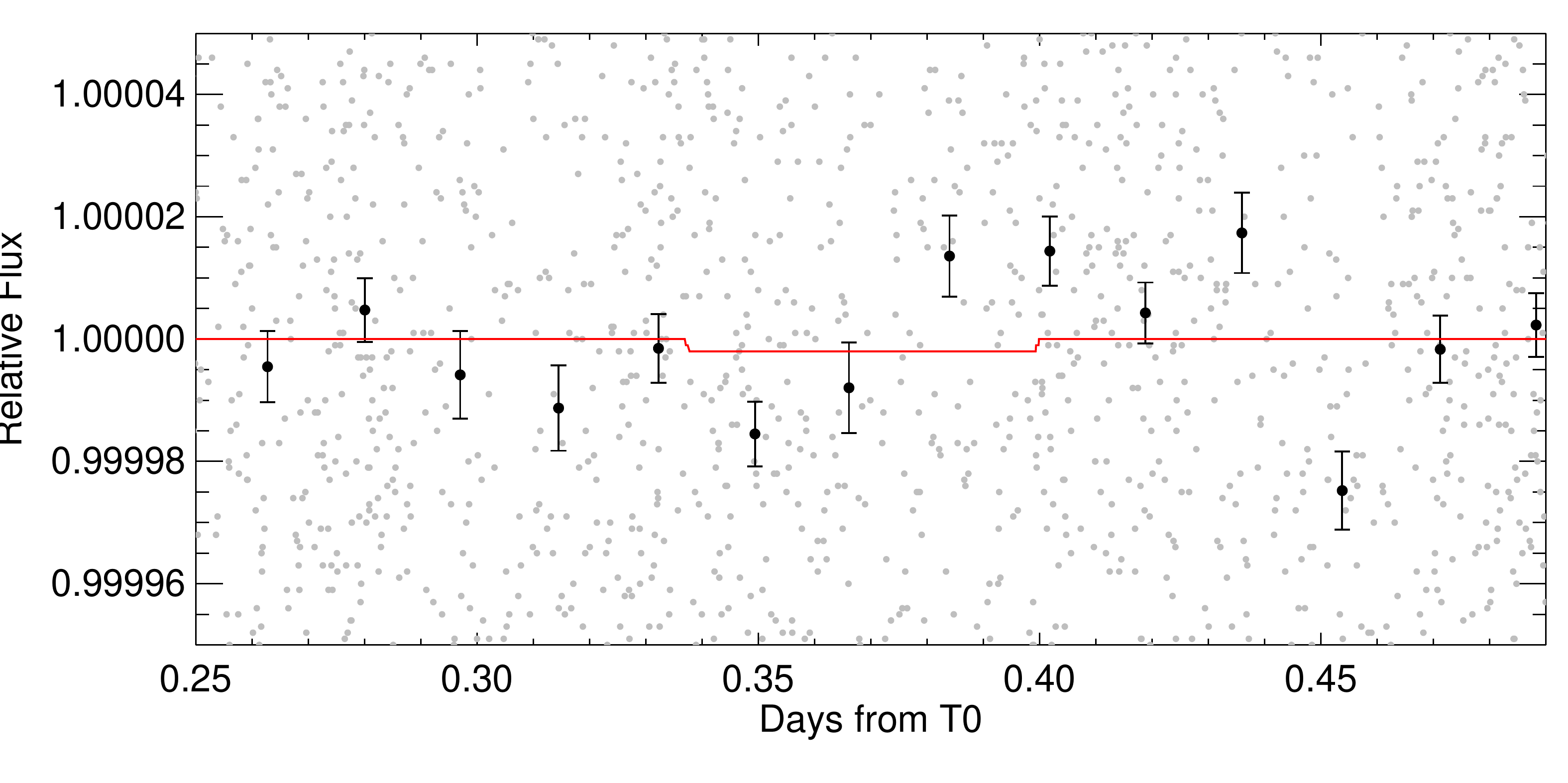}
    \caption{CHEOPS phase-folded occultation of 55\,Cnc\,e for each season: 2020 (top, two epochs), 2021 (middle, 28 epochs) and 2022 (bottom, 11 epochs). Grey circles are un-binned data points while 15-minute bins are shown as black circles. The best-fit model is shown in red.}
    \label{fig:seasons}
\end{figure}

The 2020-2022 phase-folded occultation light curve shown in Fig.~\ref{fig:lc} exhibits significant correlated noise. Most notably, the occultation duration appears $\sim$30\% shorter than the model and a trough of a similar depth is noticeable 2 hours after the occultation centre (0.46 days from $T_0$). Inspecting the inter-seasonal phase-folded light curves in Fig.~\ref{fig:seasons} reveals that this flux dip appears both in 2021 and 2022, and remains whether we use the PIPE imagette or DRP reductions. The average occultation depth of the two events in 2020 is -12$\pm$18 ppm, while it is 14$\pm$4 ppm for the 28 events in 2021, and 3$\pm$5 ppm for the 11 events in 2022.

\section{Subset of individual CHEOPS 55\,Cnc\,e occultations}
\label{app:seldepths}

\begin{figure}[!ht]
    \centering
    \includegraphics[width=0.48\textwidth]{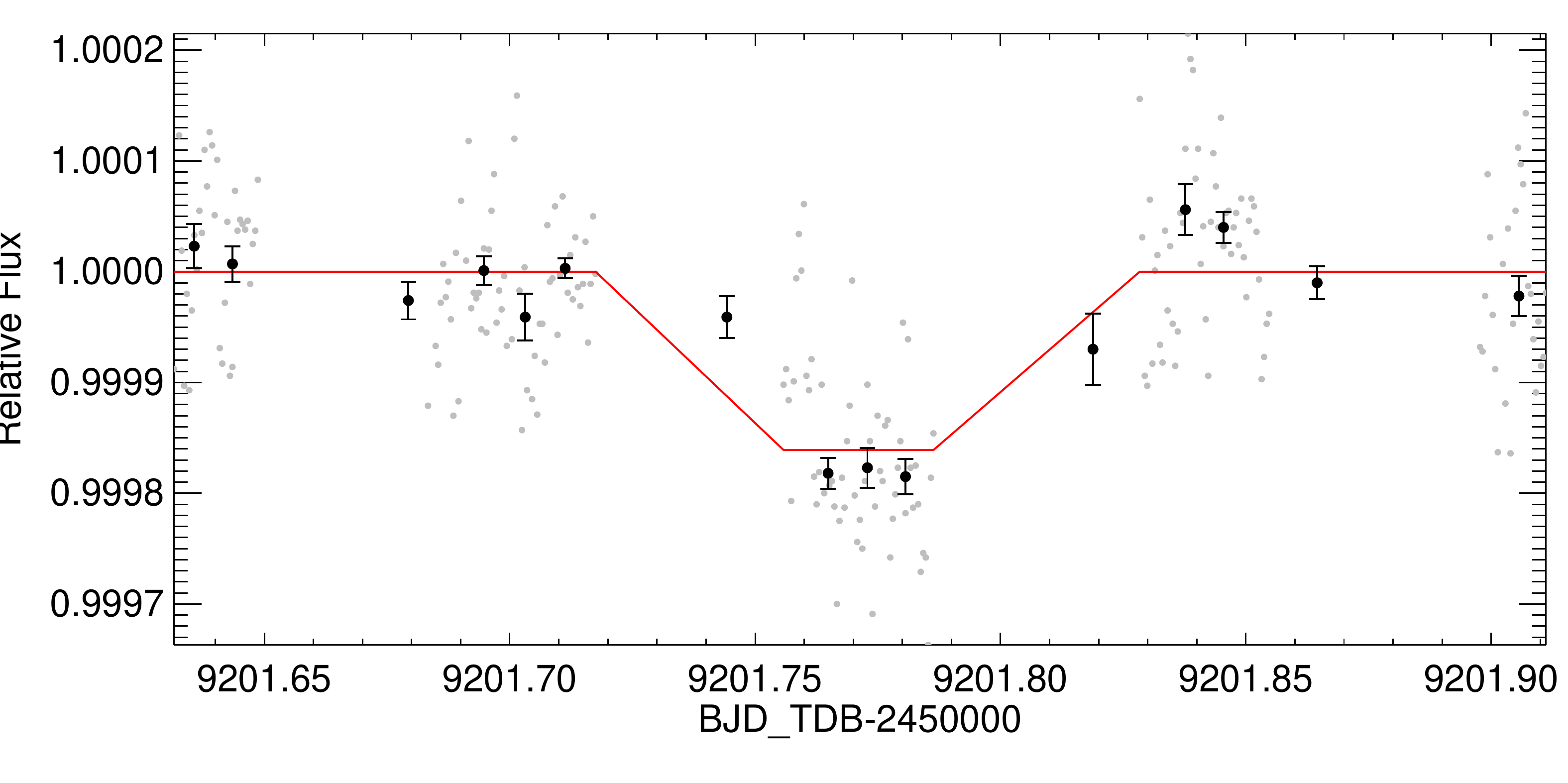}
    \includegraphics[width=0.48\textwidth]{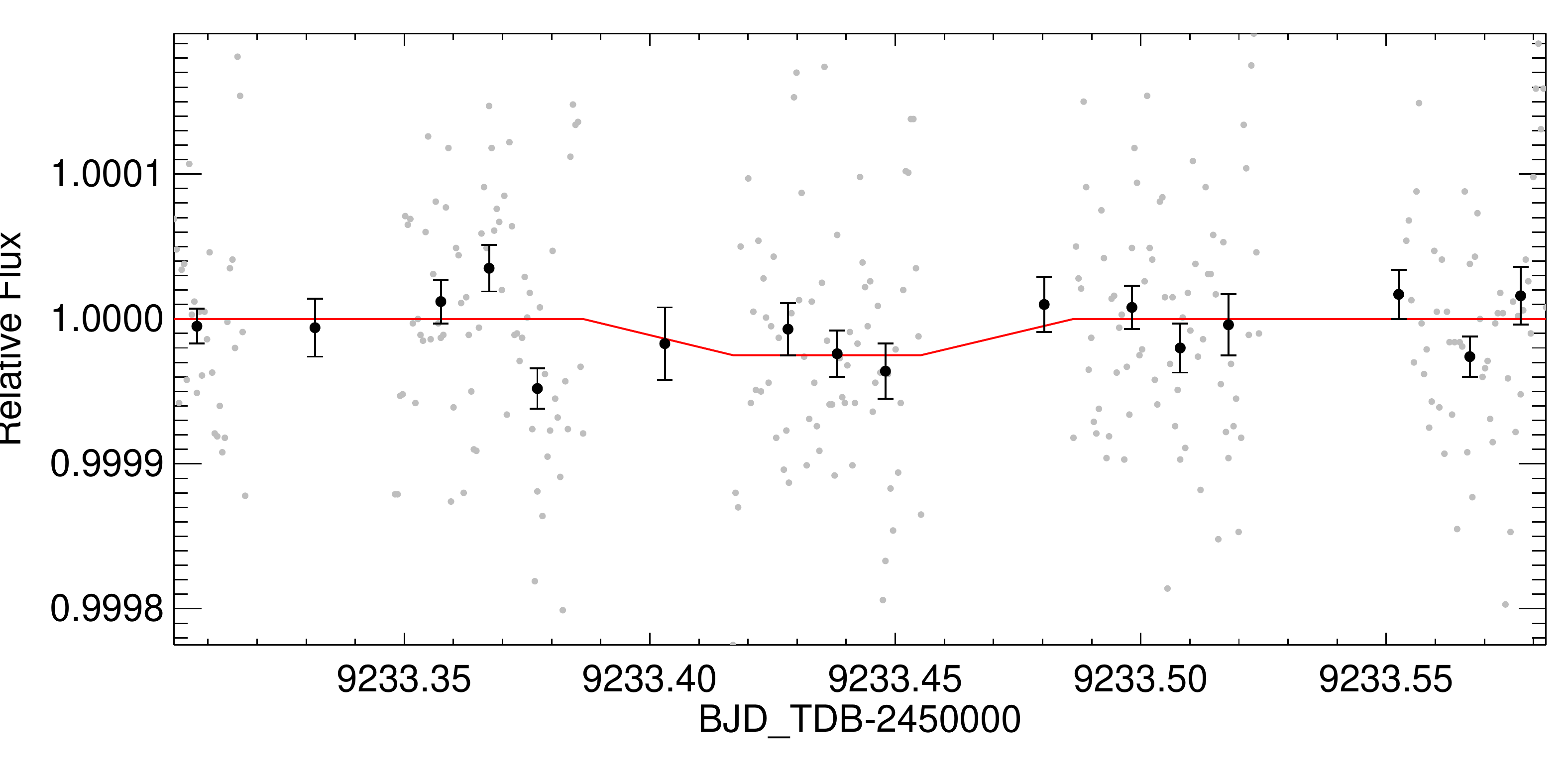}
    \includegraphics[width=0.48\textwidth]{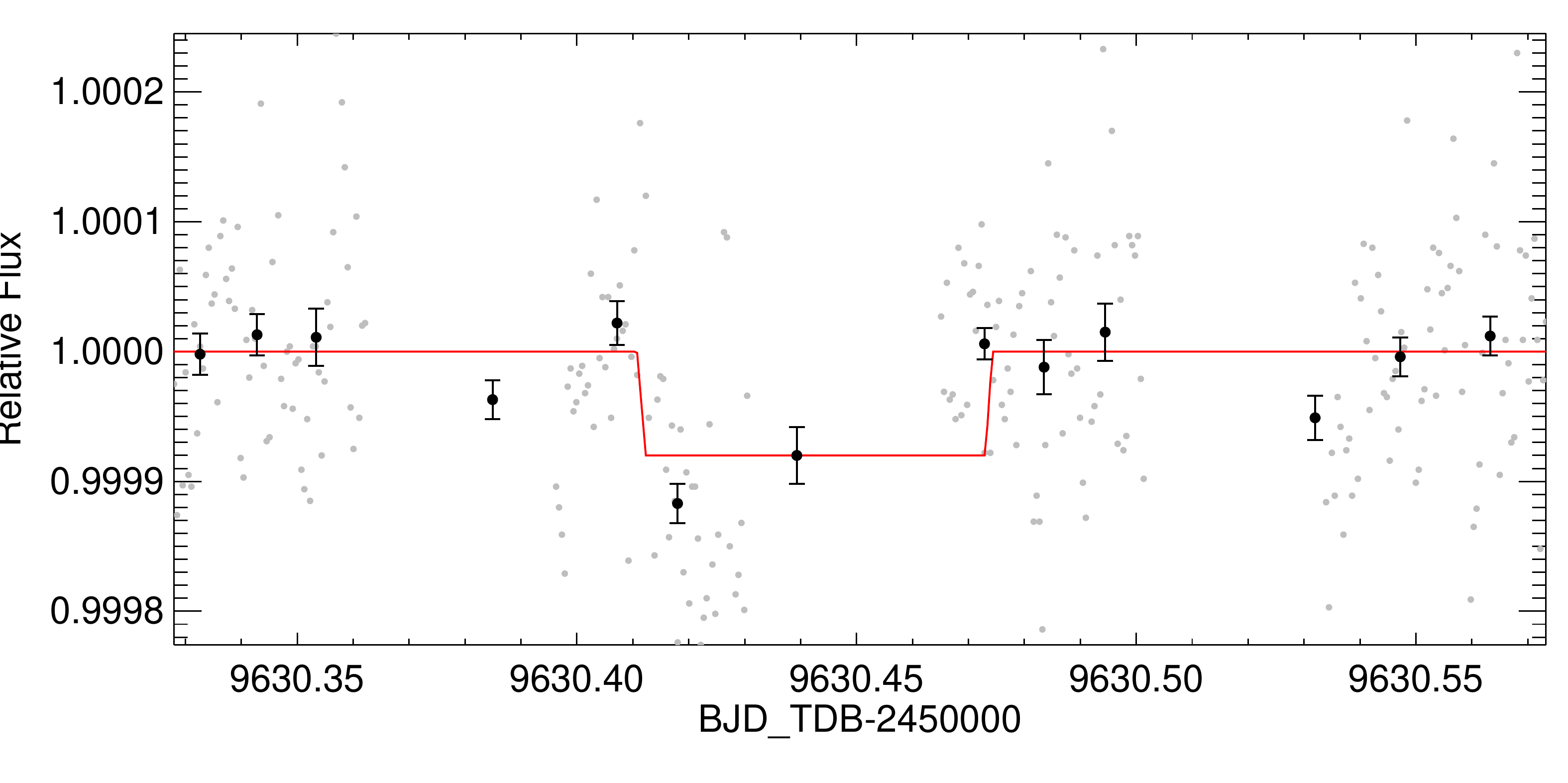}
    \includegraphics[width=0.48\textwidth]{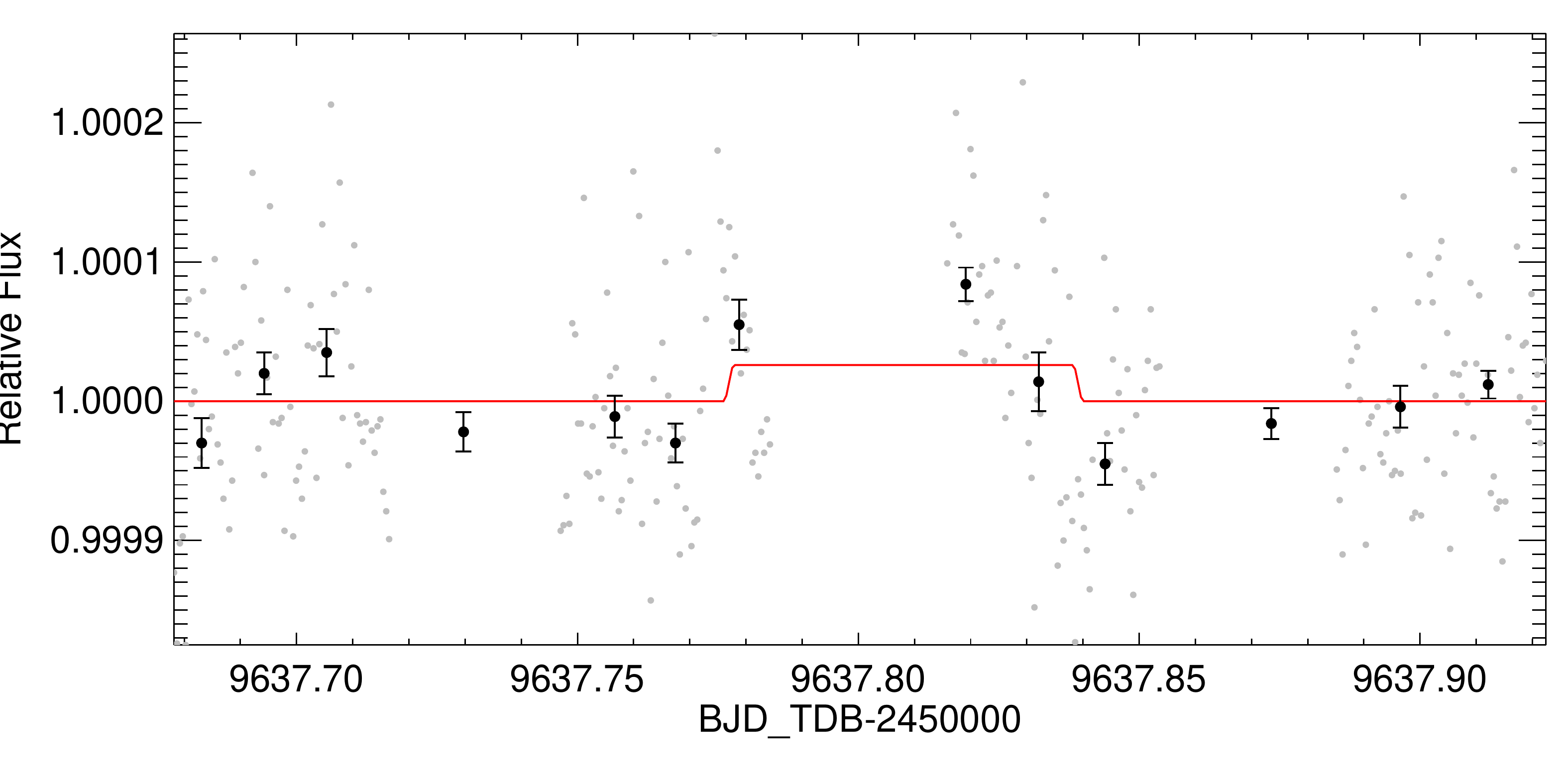}
    \caption{Selected CHEOPS individual occultations of 55\,Cnc\,e reported in Table~\ref{tab:depths} and shown in Fig.~\ref{fig:depths}. Occultation \#5 (top), Occultation \#16 (middle-top), Occultation \#38 (middle-bottom), and Occultation \#40 (bottom). All but Occultation \#16 are outliers. Grey points are un-binned time series, while 15-minute bins are shown as black points. The best-fit model is shown in red.}
    \label{fig:seldepths}
\end{figure}

\end{appendix}

\end{document}